\documentclass[usenatbib]{mn2e}
\usepackage{graphicx}

\title[Results from WASP0 II: Stellar Variability in the Pegasus Field]
{Results from the Wide Angle Search for Planets Prototype (WASP0) II:
Stellar Variability in the Pegasus Field}
\author[S. R. Kane et al.]{Stephen R. Kane$^1$, T. A. Lister$^1$,
Andrew Collier Cameron$^1$, Keith Horne$^1$,
\newauthor David James$^{2,3}$, Don L. Pollacco$^4$, Rachel A. Street$^4$,
Yiannis Tsapras$^5$\\
$^1$School of Physics \& Astronomy, University of St Andrews, North Haugh,
St Andrews, Fife KY16 9SS, Scotland\\
$^2$Department of Physics \& Astronomy, Vanderbilt University, Nashville,
TN 37235, USA\\
$^3$Laboratoire d'Astrophysique, Observatoire de Grenoble, BP 53, F-38041,
Grenoble, Cedex 9, France\\
$^4$School of Mathematics and Physics, Queen's University, Belfast,
University Road, Belfast, BT7 1NN, Northern Ireland\\
$^5$School of Mathematical Sciences, Queen Mary University of London,
Mile End Road, London, E1 4NS, UK}

\begin{document}

\maketitle

\begin{abstract}

Recent wide-field photometric surveys which target a specific field for
long durations are ideal for studying both long and short period stellar
variability. Here we report on 75 variable stars detected during
observations of a field in Pegasus using the WASP0 instrument, 73 of
which are new discoveries. The variables detected include 16 $\delta$
Scuti stars, 34 eclipsing binaries, 3 BY Draconis stars, and 4
RR Lyraes. We estimate that the fraction of stars in the field brighter
than $V \sim 13.5$ exhibiting variable behaviour with an amplitude
greater than 0.6\% rms is $\sim 0.4$\%. These results are compared with
other wide-field stellar variability surveys and implications for
detecting transits due to extra-solar planets are discussed.

\end{abstract}


\section{Introduction}

Stellar variability is a subject of great interest as there is a wide
variety of object classes producing a range of lightcurve forms. Interest
in variable stars has increased in recent years, due in no small part to
large-scale surveys and the capabilities to reduce vast datasets. There are
a number of all-sky surveys (e.g., \citet{woz04}) whose data have
identified many previously unknown variables. However, these surveys are
not always suitable for detecting short-period low-amplitude variability
due to the infrequent sampling time. A large number of variable star
detections have also resulted from intensive monitoring programs. Perhaps
the largest datasets to emerge from photometric monitoring surveys are from
microlensing experiments such as MACHO, OGLE, and EROS-2, whose
observations generally target the Galactic bulge and satellite galaxies.
Many variable star discoveries from these groups have already been
published (e.g., \citet{alc03,woz02,der02}) with much data still to be
mined.

With the recent activity surrounding detection of transiting extra-solar
planets, additional monitoring of selected fields has been undertaken.
These surveys are generally divided into cluster transit searches (e.g.,
\citet{moc02,str03a}) and wide-field transit searches
(e.g., \citet{bor01,kan04}).
As well as producing candidate extra-solar planet detections, these
surveys have yielded a wealth of new additions to variable star
catalogues. In particular, transit surveys of stellar clusters have
yielded the discovery of new variable stars as a result of their
observations (e.g., \citet{moc02,str02}). Wide-field observations of
field stars have also produced a number of new variable stars (e.g.,
\citet{bak02,eve02,har04}).

The Wide Angle Search for Planets prototype (hereafter WASP0) is a
wide-field (9-degree) instrument mounted piggy-back on a commercial
telescope and is primarily used to search for planetary transits.
WASP0 has been used to monitor three fields at two separate sites in 2000
and 2002. The monitoring programs undertaken using WASP0 make the data an
excellent source for both long and short period variability detection.

We present results from a monitoring program which targeted a field in
Pegasus. The field was monitored in order to test the capabilities of
WASP0 by detecting the known transiting planet around HD 209458. As a
by-product of this test, lightcurves of thousands of additional stars in
the field were also produced and a number of new variable stars have been
found. Here we report on 75 variable stars detected, of which 73 are new
discoveries. We estimate the fraction of stars in the field exhibiting
variable behaviour and make a comparison with other such studies of
stellar variability amongst field stars. We conclude with a discussion of
implications for extra-solar planetary transit surveys.

\section{Observations}

The WASP0 instrument is an inexpensive prototype whose primary aim is to
detect transiting extra-solar planets. The instrument consists of a
6.3cm aperture F/2.8 Nikon camera lens, Apogee 10 CCD detector (2K
$\times$ 2K chip, 16-arcsec pixels) which was built by Don Pollacco at
Queen's University, Belfast. Calibration frames were used to measure the
gain and readout noise of the chip and were found to be 15.44 e$^-$/ADU
and 1.38 ADU respectively. Images from the camera are digitized with
14-bit precision giving a data range of 0--16383 ADUs. The instrument
uses a clear filter which has a slightly higher red transmission than
blue.

The first observing run of WASP0 took place on La Palma, Canary Islands
during 2000 June 20 -- 2000 August 20. During its observing run on La
Palma, Canary Islands, WASP0 was mounted piggy-back on a commercial 8-inch
Celestron telescope with a German equatorial mount. Observations on La
Palma concentrated on a field in Draco which was regularly monitored for
two months. These Draco field observations were interrupted on four
occasions when a planetary transit of HD 209458 was predicted. On those
nights, a large percentage of time was devoted to observing the HD 209458
field in Pegasus. Exposure times for the four nights were 5, 30, 50, and
50 seconds respectively. The data for each night were rebinned into 60
second frames for ease of comparison. This campaign resulted in the
successful detection of the planet transiting HD 209458, described in
more detail in \citet{kan04}.

\section{Data Reduction}

The reduction of the WASP0 data proved to be a challenging task as
wide-field images contain many spatially dependent aspects, such as the
airmass and the heliocentric time correction. The most serious issues
arise from vignetting and barrel distortion produced by the camera optics
which alter the position and shape of stellar profiles. The data reduction
pipeline which has been developed and tested on these data is able to
solve many of these problems to produce high-precision photometry.

The pipeline first automatically classifies frames by statistical
measurements and the frames are classified as one of bias, flat, dark,
image, or unknown. A flux-weighted astrometric fit is then performed on
each image frame through cross-identification of stars with objects in the
Tycho-2 \citep{hog00} and USNO-B \citep{mon03} catalogues. This produces an
output catalogue which is ready for the photometry stage. Rather than fit
the variable point-spread function (PSF) shape of the stellar images,
weighted aperture photometry is used to compute the flux centred on
catalogue positions. The resulting photometry still contains time and
position dependent trends which we removed by post-photometry calibration.

Post-photometry calibration of the data is achieved through the use of
code which constructs a theoretical model. This model is subtracted from
the data leaving residual lightcurves which are then fitted via an
iterative process to find systematic correlations in the data. This
process also allows the separation of variable stars from the bulk of the
data since the $\mathrm{rms} / \sigma$ for variables will normally be
significantly higher than that for relatively constant stars, depending
upon the amplitude and period of the variability. The reduction of the
WASP0 data is described in more detail in \citet{kan04}.

\section{Variable Star Detection}

In this section we describe the methods used for sifting the variable stars
from the data. We then discuss the classification of the variable stars and
the methods applied for period determination.

\subsection{Photometric Accuracy}

\begin{figure}
  \includegraphics[width=8.2cm]{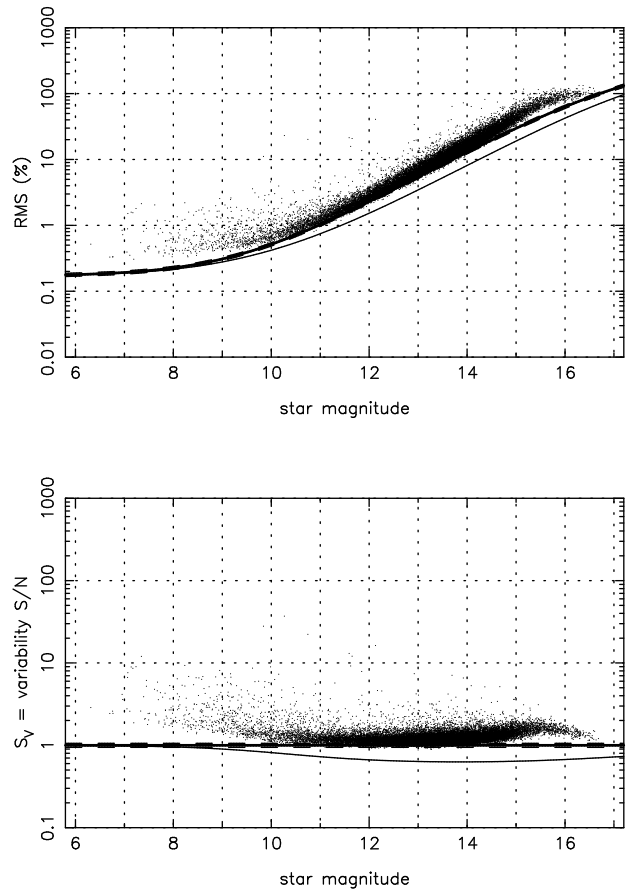}
  \caption{Photometric accuracy versus magnitude diagram including 22000
stars from one night of WASP0 observations. The upper panel shows the rms
accuracy in magnitudes in comparison with the theoretical accuracy
predicted based on the CCD noise model. The lower panel is the ratio of
the observed rms divided by the predicted accuracy. Around 4\% of stars
have an rms better than 1\%.}
\end{figure}

\begin{figure*}
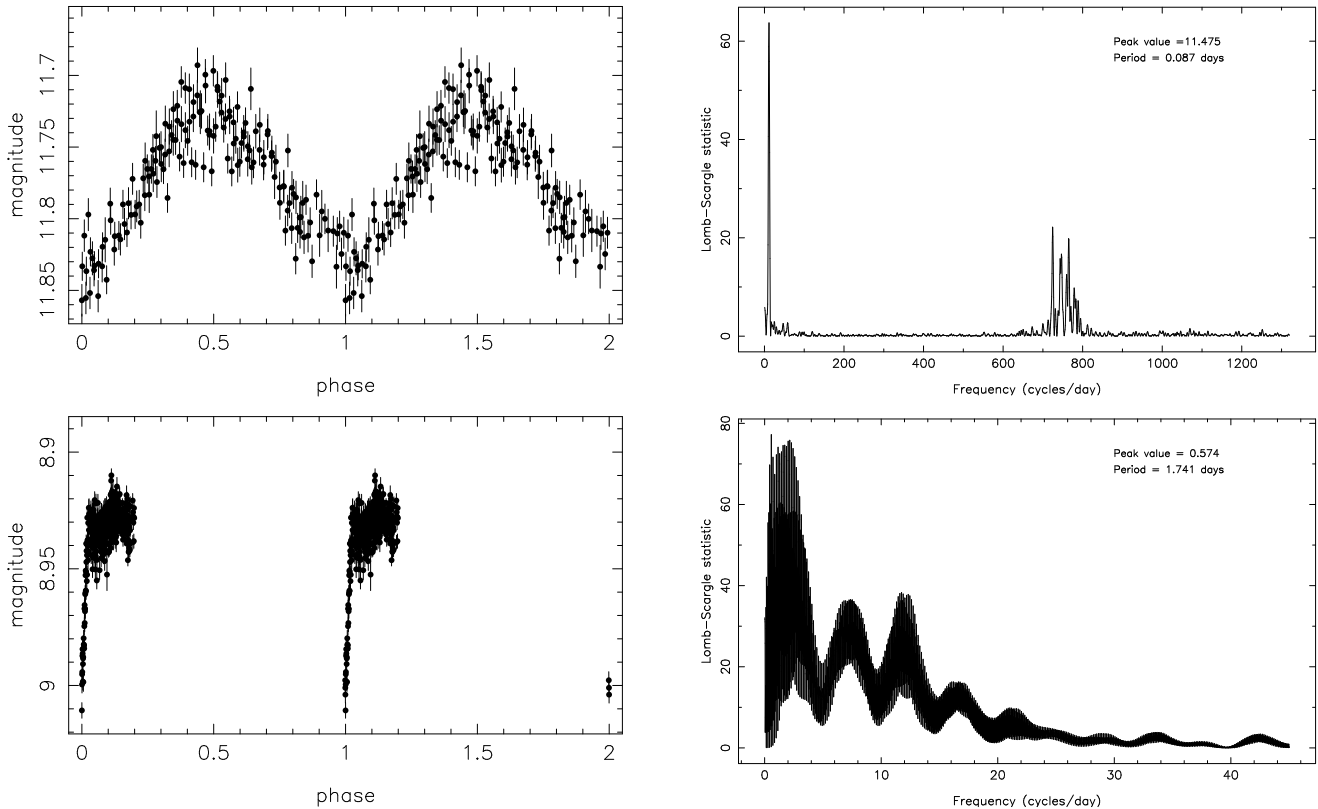

  \begin{center}
    \begin{tabular}{cc}
      \includegraphics[angle=270,width=8.2cm]{figure02a.ps} &
      \hspace{0.5cm}
      \includegraphics[angle=270,width=8.2cm]{figure02b.ps} \\
      \includegraphics[angle=270,width=8.2cm]{figure02c.ps} &
      \hspace{0.5cm}
      \includegraphics[angle=270,width=8.2cm]{figure02d.ps} \\
    \end{tabular}
  \end{center}
  \caption{Folded lightcurve (left) and periodogram (right) for two of the
variable stars detected in the Pegasus field.}
\end{figure*}

Obtaining adequate photometric accuracy is one of the many challenges
facing wide-field survey projects such as WASP0. This has been overcome
using the previously described pipeline, the results of which are
represented in Figure 1. The upper panel shows the rms versus magnitude
diagram and the lower panel shows the same rms accuracy divided by the
predicted accuracy for the CCD. The data shown include around 22000 stars
at 311 epochs from a single night of WASP0 observations and only includes
unblended stars for which a measurement was obtained at $> 20$\% of epochs.
The upper curve in each diagram
indicates the theoretical noise limit for aperture photometry with the
1-$\sigma$ errors being shown by the dashed lines either side. The lower
curve indicates the theoretical noise limit for optimal extraction using
PSF fitting.

The first stage of mining variable stars from the data was performed by a
combination of visual inspection and selecting those stars with the
highest rms/$\sigma$. Visual inspection of the stars was made far easier by
the dense sampling of the field which would otherwise have obscured many
short-period variables in the data. The selected stars were then
extracted and analysed using a spectral analysis technique which will now
be described in greater detail.

\subsection{Model Fitting and Period Calculation}

In analysing variable stars, there are various methods that can be used
to extract period information from the lightcurves. For this analysis, we
make use of the ``Lomb method'' \citep{pre92} which is especially suited
to unevenly sampled data. This method uses the Nyquist frequency to
perform spectral analysis of the data resulting in a Lomb-Scargle
statistic for a range of frequencies. The Lomb-Scargle statistic indicates
the significance level of the fit at that particular frequency and hence
yields the likely value for the period. This method generally works quite
well for data which are a combination of sines and cosines, but care must
be taken with non-sinusoidal data as the fitted period may be half or
twice the value of the true period.

Fortran code was written to automatically apply the Lomb method to each
of the suspected variable star lightcurves and attempt to determine the
period, which could then be used to produce phase-folded lightcurves. The
stars are then sorted according to the fitted period and examined
individually. Figure 2 shows the periodogram and the folded lightcurve for
two of the variable stars. The upper lightcurve has good phase coverage
and a sinusoidal shape whilst the lower lightcurve has poor phase coverage
and only a single eclipse was observed.

Data were acquired for $\sim 6$ hours per night for a total of 4 nights.
Each of these nights were spaced 7 nights apart. The implications of this
are that the data are far more sensitive to short ($< 0.5$ day) period
variables, but also that we can expect a bias towards shorter periods due
to the large gaps between observations which produces significant multiple
peaks in the associated periodograms. In the case of stars for which only
one photometric variation is observed over the entire observing run, the
bias is towards longer periods since the spectral analysis is not
constrained by the data from the other nights. The lower lightcurve shown
in Figure 2 is an example of this effect as demonstrated by the strong
aliasing visible in the periodogram.

\section{Results}

\begin{figure*}
  \includegraphics[angle=270,width=16.0cm]{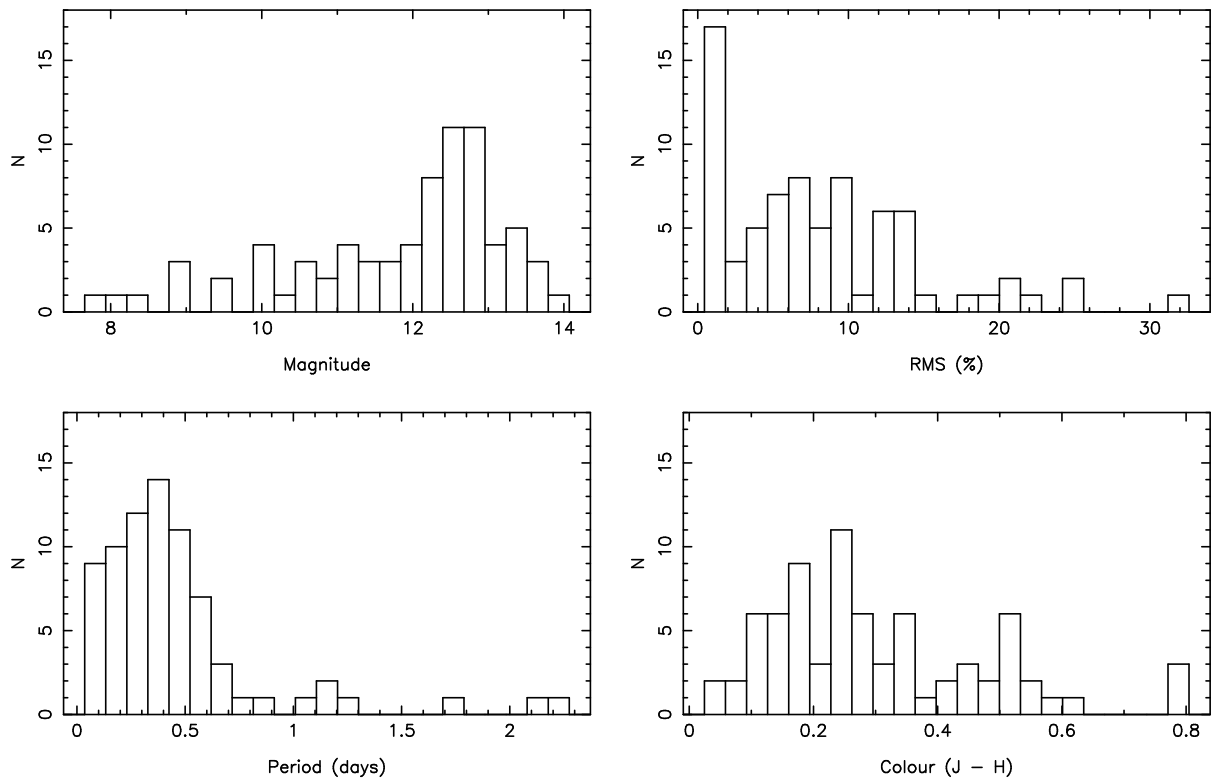}
  \caption{Histogram of variable star magnitudes (top-left), rms
(top-right), periods (bottom-left), and colour (bottom-right).}
\end{figure*}

In total, 75 variable stars were identified in the WASP0 Pegasus data,
as listed in Table 1 in order of increasing period.
Through the use of the databases SIMBAD \citep{wen00} and VIZIER
\citep*{och00}, it was found that 73 are previously unknown variables.
To assist in the classification of the variables, colour information
provided by the Two Micron All Sky Survey (2MASS) project was utilised.
This provided accurate colours using $J$, $H$, and $K$ filters down to the
magnitude limits of the WASP0 data.
Shown in Figure 3 are histograms for the instrumental magnitude, rms,
period, and colour for the detected variable stars. Of particular interest
are the significant number of small rms (and hence small amplitude)
variables detected. The period distribution peaks at slightly less than
0.5 days as expected.

\begin{table*}
\caption{List of detected variable stars sorted in order of increasing
period. Known variable stars are marked with a *. Variables for which there
is an associated question mark have a small uncertainty regarding the
classification. Variables for which a suitable classification could not be
assigned are designated as unknown (U).}
\begin{tabular}{@{}lccccccc}
star \# & catalogue \# & period (days) & mag & rms (\%) & $J - H$ &
$H - K$ & class\\
 1     & Tycho 1689-00007-1 & 0.049 & 10.854 &  1.788 & 0.06 & 0.03 & $\delta$ Scuti    \\
 2     & Tycho 1682-00151-1 & 0.070 &  8.344 &  0.938 & 0.09 & 0.01 & $\delta$ Scuti    \\
 3     & Tycho 1691-01562-1 & 0.081 & 12.935 &  7.051 & 0.20 & 0.04 & $\delta$ Scuti    \\
 4     & Tycho 1682-00125-1 & 0.087 & 11.770 &  3.879 & 0.05 & 0.06 & $\delta$ Scuti    \\
 5     & Tycho 2202-01669-1 & 0.094 & 12.787 &  9.129 & 0.11 & 0.05 & $\delta$ Scuti    \\
 6     & Tycho 1683-00394-1 & 0.102 & 10.039 &  0.660 & 0.10 & 0.07 & $\delta$ Scuti    \\
 7     & USNO  1077-0716280 & 0.113 & 13.374 &  9.357 & 0.15 & 0.00 & $\delta$ Scuti    \\
 8     & Tycho 1674-00459-1 & 0.120 & 10.547 &  1.629 & 0.14 & 0.04 & $\delta$ Scuti    \\
 9     & USNO  1080-0687717 & 0.122 & 12.388 &  2.870 & 0.62 & 0.16 & $\delta$ Scuti    \\
10     & Tycho 1693-02012-1 & 0.135 &  9.386 &  0.630 & 0.17 &-0.25 & $\delta$ Scuti    \\
11     & USNO  1083-0635832 & 0.135 & 12.621 &  6.928 & 0.33 & 0.08 & $\delta$ Scuti    \\
12     & Tycho 1683-01839-1 & 0.153 &  9.935 &  1.690 & 0.12 & 0.05 & $\delta$ Scuti    \\
13     & USNO  1063-0610342 & 0.168 & 13.537 & 19.764 & 0.16 & 0.05 & $\delta$ Scuti    \\
14     & Tycho 1681-01275-1 & 0.200 & 12.278 &  5.764 & 0.26 & 0.05 & $\delta$ Scuti    \\
15     & USNO  1098-0571100 & 0.203 & 11.054 &  1.133 & 0.52 & 0.06 & U     \\
16     & USNO  1082-0661701 & 0.203 & 12.640 &  4.997 & 0.51 & 0.04 & U     \\
17     & USNO  1079-0712902 & 0.207 & 12.556 &  8.334 & 0.49 & 0.08 & U     \\
18     & Tycho 1685-01214-1 & 0.217 & 11.361 &  2.566 & 0.48 & 0.12 & U     \\
19     & USNO  1062-0603571 & 0.230 & 13.287 & 13.675 & 0.25 & 0.00 & U     \\
20     & USNO  1062-0604608 & 0.259 & 12.707 &  8.872 & 0.34 & 0.08 & U     \\
21     & USNO  1118-0583493 & 0.261 & 13.330 & 12.253 & 0.40 & 0.04 & U     \\
22     & USNO  1109-0567069 & 0.271 & 13.559 &  9.350 & 0.44 & 0.08 & EW    \\
23     & USNO  1084-0602522 & 0.276 & 12.736 &  8.405 & 0.50 & 0.09 & EW    \\
24     & USNO  1088-0594050 & 0.279 & 12.904 & 10.692 & 0.23 & 0.08 & U     \\
25     & USNO  1051-0620188 & 0.291 & 12.287 &  6.450 & 0.26 & 0.08 & U     \\
26     & Tycho 1690-01643-1 & 0.291 &  9.916 &  1.536 & 0.24 &-0.11 & EW    \\
27     & USNO  1130-0634392 & 0.292 & 12.419 &  8.573 & 0.45 & 0.17 & EW    \\
28     & USNO  1083-0636244 & 0.297 & 14.033 & 24.659 & 0.36 & 0.20 & EW    \\
29     & USNO  1077-0715703 & 0.309 & 12.264 &  8.284 & 0.60 & 0.20 & EW?   \\
30     & USNO  1099-0576195 & 0.310 & 12.127 &  7.187 & 0.33 & 0.04 & EW    \\
31     & USNO  1067-0619020 & 0.318 & 12.721 &  9.156 & 0.42 & 0.08 & U     \\
32     & Tycho 1687-01479-1 & 0.342 & 12.051 &  4.822 & 0.21 & 0.08 & EW    \\
33     & USNO  1124-0632100 & 0.346 & 13.264 & 24.698 & 0.46 & 0.14 & EW    \\
34     & Tycho 1670-00251-1 & 0.347 & 11.851 & 14.219 & 0.35 & 0.09 & EW    \\
35     & USNO  1077-0715670 & 0.349 & 13.368 & 12.316 & 0.52 & 0.12 & EW?   \\
36     & USNO  1085-0593094 & 0.349 & 11.462 &  1.610 & 0.25 & 0.07 & EW?   \\
37     & USNO  1123-0641431 & 0.370 & 13.006 &  9.257 & 0.28 & 0.07 & EW?   \\
38     & Tycho 1666-00301-1 & 0.380 & 11.595 &  3.978 & 0.27 & 0.07 & EW?   \\
39     & Tycho 1147-01237-1 & 0.389 & 11.848 &  4.550 & 0.18 & 0.09 & EW    \\
40     & USNO  1111-0568829 & 0.399 & 12.524 & 14.847 & 0.15 & 0.06 & RR Lyrae    \\
41     & USNO  1120-0611140 & 0.406 & 12.335 &  6.101 & 0.27 & 0.04 & EW?   \\
42     & USNO  1077-0719427 & 0.407 & 12.906 &  5.343 & 0.34 & 0.04 & EW    \\
43     & USNO  1074-0692109 & 0.407 & 12.539 &  5.418 & 0.32 & 0.05 & U     \\
44     & USNO  1048-0613132 & 0.414 & 12.421 &  8.323 & 0.23 & 0.08 & U     \\
45     & Tycho 1684-01512-1 & 0.422 & 10.225 &  0.659 & 0.17 & 0.11 & U     \\
46     & USNO  1096-0579000 & 0.433 & 13.763 & 13.391 & 0.56 & 0.17 & EW?   \\
47     & Tycho 2203-01663-1 & 0.435 & 10.080 & 12.856 & 0.21 & 0.01 & EW    \\
48     & USNO  1127-0652941 & 0.440 & 13.071 & 18.065 & 0.25 & 0.08 & EW    \\
49     & USNO  1067-0619189 & 0.443 & 12.500 &  5.927 & 0.34 &-0.01 & U     \\
50     & Tycho 1666-00208-1 & 0.453 &  8.952 &  0.740 & 0.80 & 0.26 & BY Draconis    \\
51     & Tycho 1684-00522-1 & 0.456 & 12.119 &  9.499 & 0.28 & 0.06 & EW    \\
52     & Tycho 1687-00659-1 & 0.487 & 10.515 & 20.782 & 0.23 & 0.05 & EB    \\
53     & Tycho 1685-01784-1 & 0.488 & 12.373 & 32.408 & 0.14 & 0.09 & RR Lyrae    \\
54$^*$ & Tycho 2202-01379-1 & 0.502 & 11.057 & 21.161 & 0.24 & 0.06 & RR Lyrae    \\
55     & USNO  1063-0601782 & 0.506 & 12.902 &  6.996 & 0.29 & 0.01 & U     \\
56     & Tycho 1688-01026-1 & 0.515 & 11.856 & 21.775 & 0.24 & 0.09 & EB    \\
57     & USNO  1126-0625388 & 0.524 & 11.173 &  5.835 & 0.80 & 0.30 & EA    \\
58     & Tycho 1685-00588-1 & 0.556 & 12.869 & 13.129 & 0.16 & 0.05 & RR Lyrae   \\
59     & USNO  1061-0601686 & 0.561 & 12.929 &  7.108 & 0.27 & 0.10 & EW?   \\
60     & USNO  1090-0579466 & 0.566 & 12.878 & 11.671 & 0.12 & 0.05 & EW?   \\
61     & Tycho 1683-00877-1 & 0.587 & 11.404 &  4.362 & 0.53 & 0.24 & EW    \\
62     & Tycho 1684-00288-1 & 0.596 & 11.711 &  4.571 & 0.17 & 0.10 & EW    \\
63     & USNO  1047-0628425 & 0.603 & 13.076 & 13.307 & 0.17 & 0.08 & EW    \\
\end{tabular}
\end{table*}

\begin{table*}
\contcaption{}
\begin{tabular}{@{}lccccccc}
star \# & catalogue \# & period (days) & mag & rms (\%) & $J - H$ &
$H - K$ & class\\
64     & Tycho 1686-00904-1 & 0.647 & 13.072 & 12.663 & 0.12 & 0.03 & EW    \\
65     & Tycho 1679-01714-1 & 0.668 &  8.195 &  0.884 & 0.13 & 0.01 & U     \\
66     & Tycho 1684-00561-1 & 0.717 & 11.085 &  2.281 & 0.03 & 0.09 & $\delta$ Scuti?   \\
67     & Tycho 1686-00469-1 & 0.737 & 10.504 & 12.776 & 0.51 & 0.07 & BY Draconis?   \\
68     & Tycho 1684-00023-1 & 0.830 & 12.605 &  6.648 & 0.39 & 0.07 & E?    \\
69$^*$ & Tycho 1674-00732-1 & 1.041 &  7.693 &  1.614 & 0.10 & 0.05 & $\delta$ Scuti?   \\
70     & Tycho 1682-00761-1 & 1.141 & 10.931 &  1.754 & 0.55 & 0.12 & U     \\
71     & USNO  1123-0632849 & 1.187 & 12.638 & 13.051 & 0.35 & 0.06 & EA    \\
72     & USNO  1125-0628249 & 1.212 & 12.459 &  9.359 & 0.16 & 0.08 & E?    \\
73     & Tycho 1691-01257-1 & 1.741 &  8.935 &  1.489 & 0.18 & 0.05 & E?    \\
74     & Tycho 1666-00644-1 & 2.158 &  8.858 &  1.008 & 0.80 & 0.24 & BY Draconis?   \\
75     & Tycho 1149-00326-1 & 2.262 &  9.480 &  1.444 & 0.18 & 0.05 & U     \\
\end{tabular}
\end{table*}

\subsection{Colour-Magnitude Diagram}

The WASP0 instrument normally uses a clear filter as described in
\citet{kan04} and so no colour information is available in the WASP0
data. However, we made use of the astrometric catalogues in this by
transforming the USNO-B colours to a more standard system.
Specifically, the USNO-B colours used were second epoch IIIa-J, which
approximates as B, and second epoch IIIa-F, which approximates as R.
\citet{kid04} describes a suitable linear transformation from USNO-B
filters to the more standard Landolt system. This colour transformation
is given by:
\begin{eqnarray}
  \mathrm{B: Landolt} & = & 1.097*\mathrm{USNO(B)} - 1.216 \nonumber \\
  \mathrm{R: Landolt} & = & 1.031*\mathrm{USNO(R)} - 0.417 \nonumber
\end{eqnarray}
A linear least-squares fit to the colours computed in \citet{bes90} was
used to convert from $B - R$ to $B - V$. Using this transformation, we
are able to construct an approximate colour-magnitude diagram as shown
in Figure 4. The variable stars cover a broad range of magnitudes and
spectral types, with almost all of them falling on the main sequence.
Several of the variables are late-type giant stars and are therefore
separate from the bulk of variables shown in Figure 4 and the colour
histogram of Figure 3. The apparent colour cutoff at $B - V \approx 1$
is an artifact from the insertion of Tycho-2 objects into the USNO-B
catalogue.
Using the 2MASS colour information, the period of the variables are
plotted as a function of $J - H$ colour in Figure 5. The well-known
period-colour relationship for contact binaries \citep{rub01} is clearly
evident in the figure. These various variable groups represented in the
figure are now discussed in more detail, and the phase-folded
lightcurves of the variables are shown in Figures 6--8.

\begin{figure}
  \includegraphics[width=8.2cm]{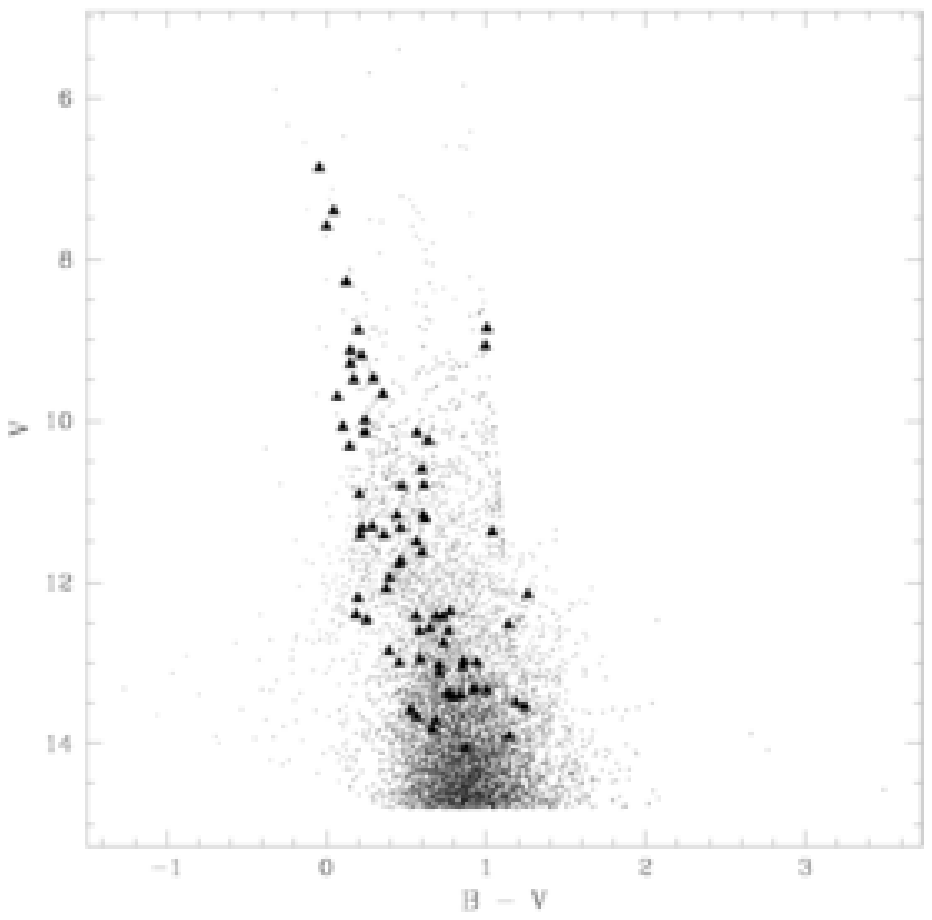}
  \caption{Colour-magnitude diagram for the Pegasus field with the
location of variable stars shown as triangles.}
\end{figure}

\begin{figure}
  \includegraphics[angle=270,width=8.2cm]{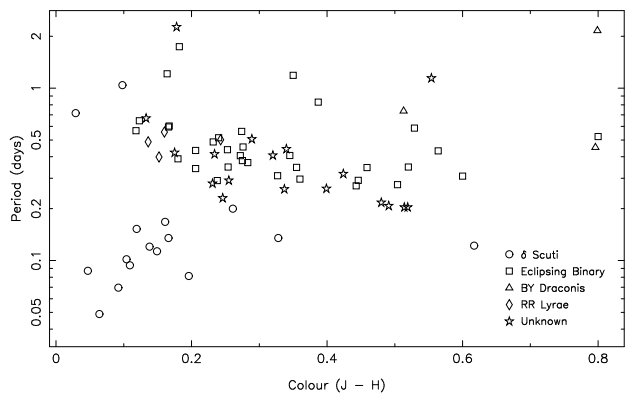}
  \caption{Log period as a function of $J - H$ colour for each of the
variable stars.}
\end{figure}

\subsection{$\delta$ Scuti Stars}

Due to the high time resolution of the WASP0 observations, the WASP0 data
are particularly sensitive to very short period variations in stellar
lightcurves. The best examples are the pulsating $\delta$ Scuti stars, of
which 15 were identified from our observations. A number of the $\delta$
Scuti stars exhibit multi-periodic behaviour, such as stars 1 and 5. Stars
66 and 69 are also of the $\delta$ Scuti type but are multi-periodic and
have been folded on the longer period. The bright star HD 207651 (star 69)
is one of the two known variables to have been detected. Observations by
\citet*{hen04} revealed this to be a triple system with the lower
frequency variations resulting from ellipsoidal variations.

\subsection{Eclipsing Binaries}

Eclipsing binaries comprise at least 45\% of the variables detected in
this survey, and most of the unclassified variables are also strongly
suspected to be eclipsing binaries. Most of the binaries are of the
W Ursae Majoris (EW) type, the remainder being possible candidates for
Algol (EA) or Beta Lyrae (EB) type binaries. The best example is star 47,
for which exceptional data quality and clear phase coverage reveal
slightly differing minima values, indicative of a close binary pair.

Of particular interest is star 57, an apparent eclipsing binary whose
colour suggests a very late-type star. Although lack of phase information
adds considerable uncertainty to the period, the estimated small period
suggests that this may be an eclipsing system with M-dwarf components.
There are only a few known eclipsing binaries of this type
\citep{mac04,rib03} and they are very interesting as they allow rare
opportunities to investigate the physical properties of late-type stars.

\subsection{BY Draconis Stars}

BY Draconis stars are generally spotted late-type stars whose variability
arises from their rotation. Late-type stars comprise the minority of our
variables list and so only 3 were identified as being of this type.

Star 74 has been classified as a BY Draconis due to the late spectral type
and variable minima of the lightcurve. However, the lack of phase
information and therefore uncertain period means that there are
undoubtedly other equally viable classifications.

\subsection{RR Lyrae Stars}

The mean period of RR Lyrae stars is around 0.5 days and are therefore
highly likely to be detected by the WASP0 observations. A total of 4
RR Lyrae stars were detected in this survey, one of which is a known
RR Lyrae (star 54), the F0 star AV Peg. The remaining three consist of
star 40 (type b), star 53 (type a), and star 58 (type c).

\begin{figure*}
  \includegraphics[angle=270,width=17.3cm]{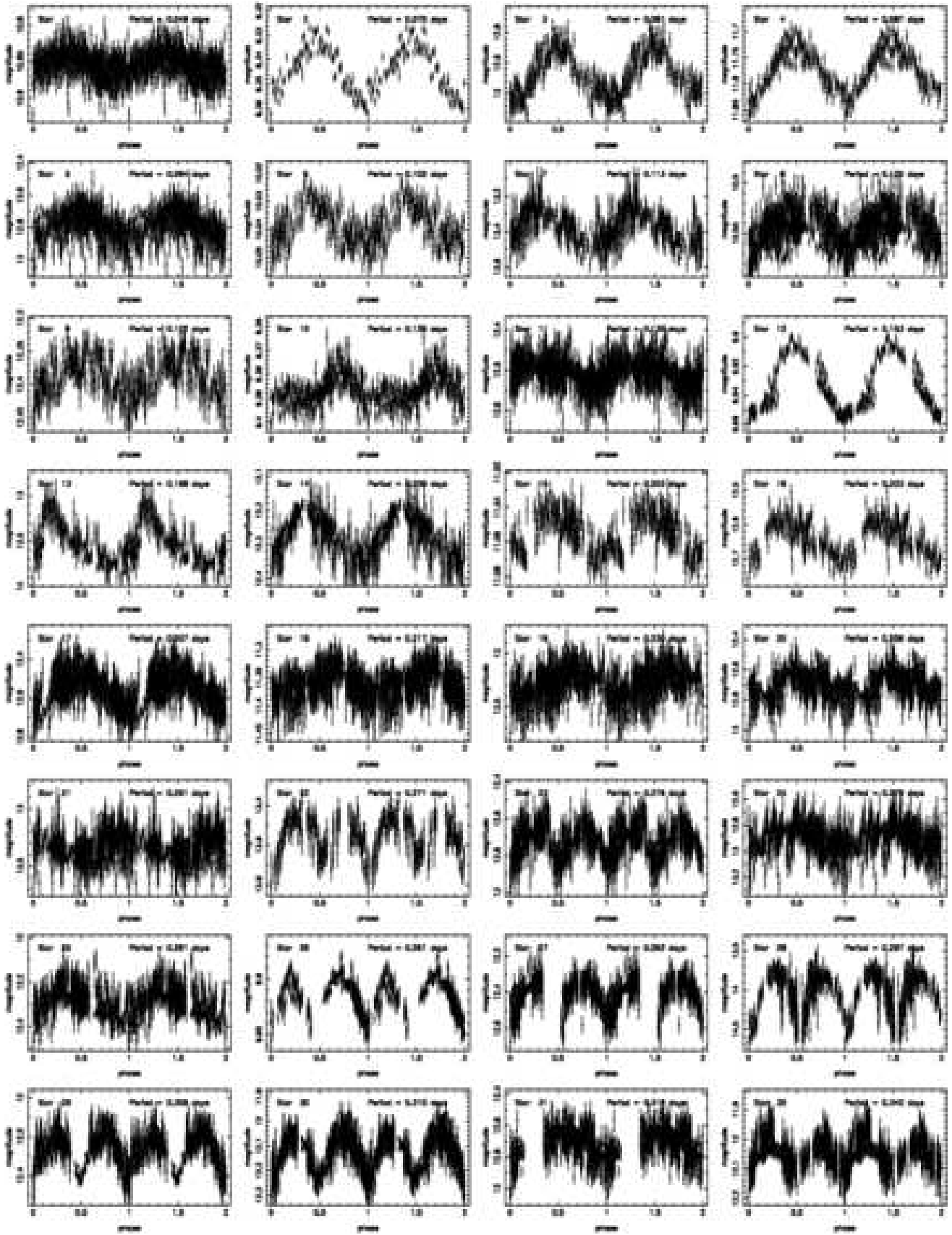}
  \caption{Variable stars 1--32 detected in the Pegasus field.}
\end{figure*}

\begin{figure*}
  \includegraphics[angle=270,width=17.3cm]{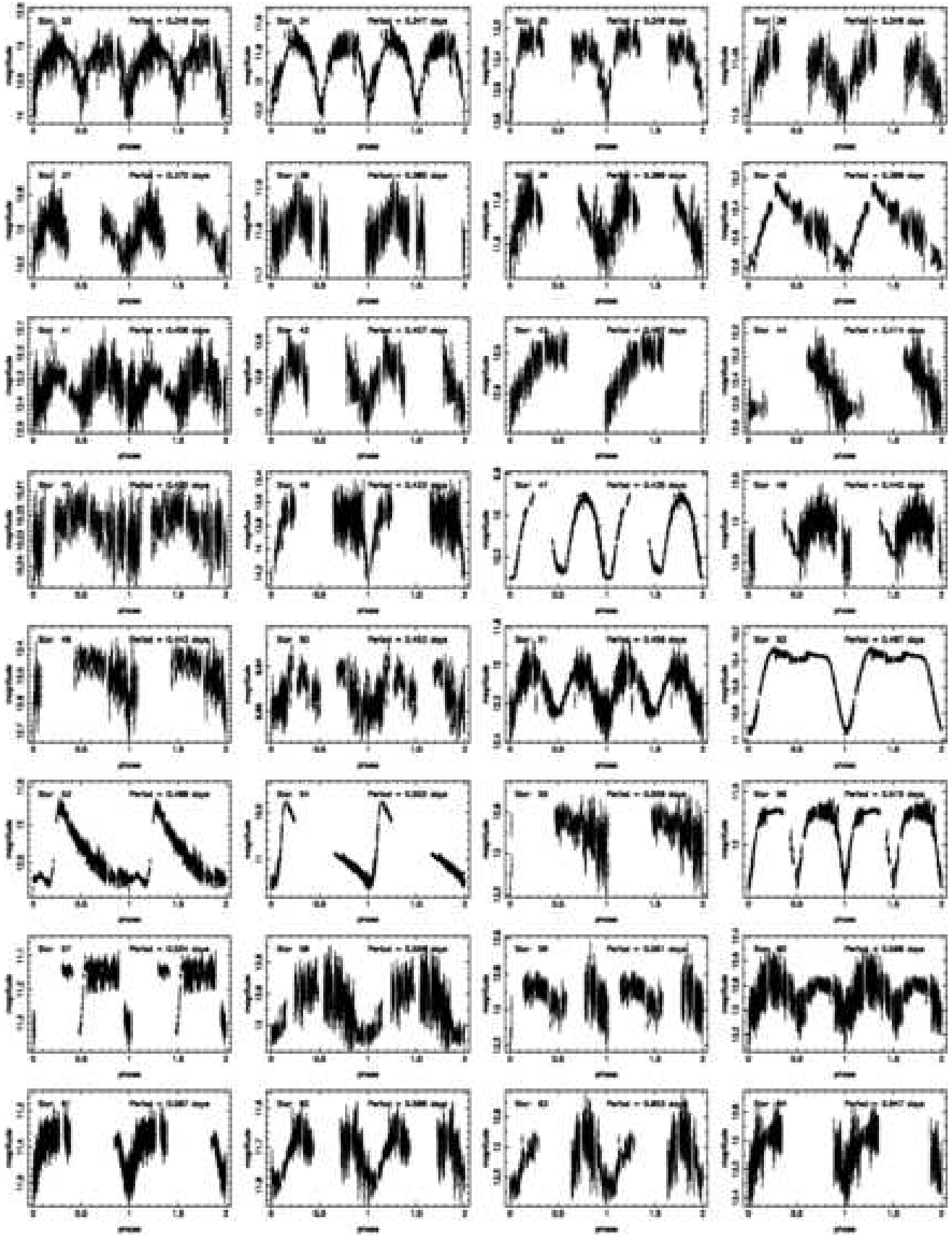}
  \caption{Variable stars 33--64 detected in the Pegasus field.}
\end{figure*}

\begin{figure*}
  \includegraphics[angle=270,width=17.3cm]{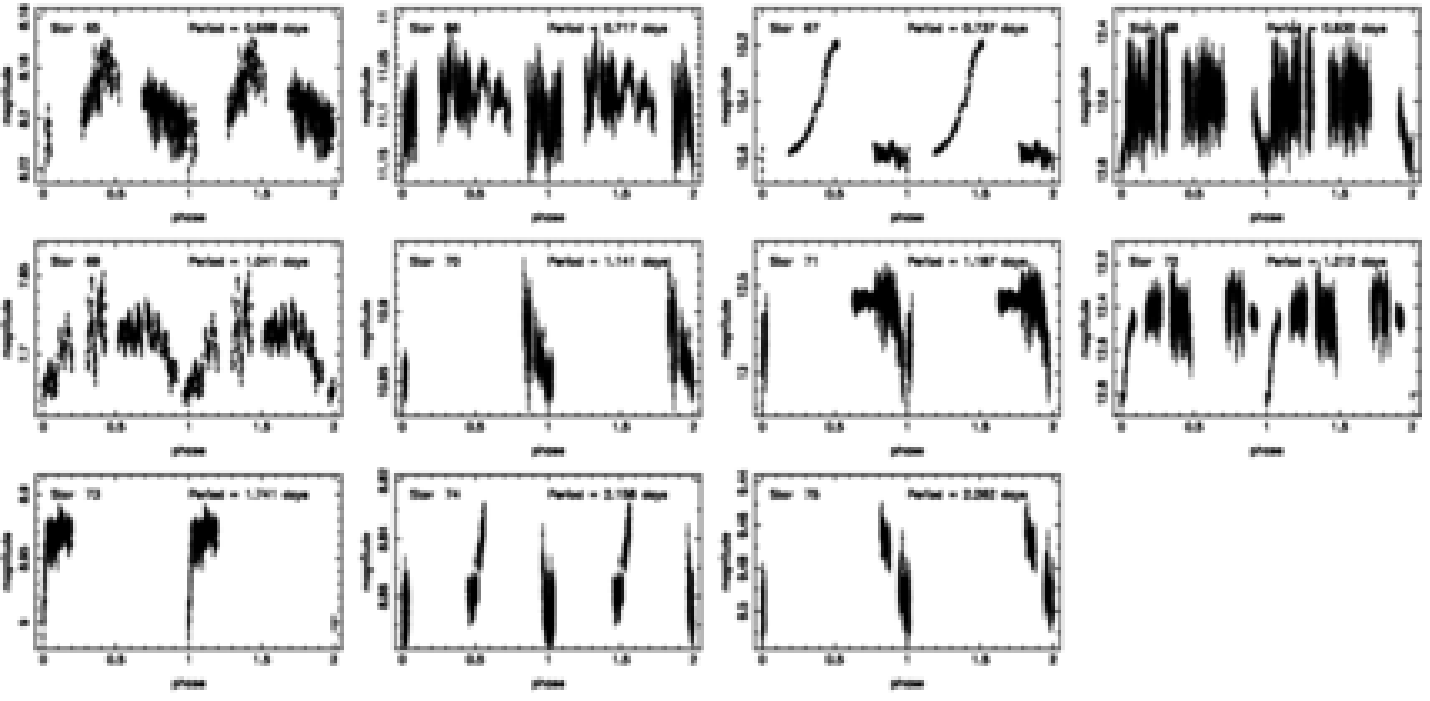}
  \caption{Variable stars 65--75 detected in the Pegasus field.}
\end{figure*}

\subsection{Unclassified and Suspected Variables}

Around 24\% of the detected variables were unable to be classified due
to a lack of S/N and/or a lack of phase information. Examples of low
S/N stars are stars 17--21. Stars 43-44 are examples of lightcurves
missing valuable phase information. For some of these stars, the stellar
image fell close to the edge of the chip and so was not observed
consistently throughout the four nights. Star 65 has the shape, period,
and colour (spectral type $\sim$ F0) of a typical RR Lyrae. However, the
small amplitude of the variability is too small and so it remained
unclassified. It is likely that many of the unclassified variables
are eclipsing binaries for which one of the eclipses was not observed,
as evidenced by their close proximity to the period-colour relationship
visible in Figure 5.

\section{Discussion}

In total, $\sim 20000$ stars were searched for variability amongst the
Pegasus field stars down to a magnitude of $\sim 13.5$. From these stars,
75 variable stars were positively identified. The techniques used to
extract the variable stars from the data detected variability to an rms
of around 0.6\%. Hence, it is estimated that $\sim 0.4$\% of Pegasus
field stars brighter than $V \sim 13.5$ are variables with an amplitude
greater than $\sim 0.6$\%. For comparison, \citet{har04} used
a similar instrument to observe a field in Cygnus and found that 
around 1.5\% of stars exhibit variable behaviour with amplitudes greater
than $\sim 3$\%. \citet{eve02} detected variability in field stars for a
much fainter magnitude range ($13.8 < V < 19.5$) and found the rate to
be as high as 17\%.
However, as discussed earlier, the results presented here are likely
to be biased towards lower periods due to the spacing of the
observations.
This also means that many long period variables may have remained
undetected. Therefore the percentage of Pegasus field stars which
exhibit variable behaviour calculated above should be considered a
lower limit, as the actual value may well be slightly higher.

The Pegasus field stars surveyed in this study are predominantly
solar-type stars. However, 60\% of the variables detected are bluer than
solar ($J - H < 0.32$) with $\sim 30$\% of these blue variables
classified as $\delta$ Scuti stars. Those variables classified as
``unknown'' are fairly evenly distributed in colour and are most likely
to be eclipsing binaries.

It has been noted by \citet{eve02} that low levels of stellar variability
are likely to interfere with surveys hunting for transits due to
extra-solar planets. Indeed it has developed into a major challenge for
transit detection algorithms to distinguish between real planetary
transits and false-alarms due to variables, especially grazing eclipsing
binaries \citep{bro03}. If many of the unclassified variable stars from
this survey are eclipsing binaries then these will account for well over
half ($> 40$) of the total variables found. This will have a significant
impact upon transit surveys, especially wide-field surveys for which the
stellar images are prone to under-sampling and therefore vulnerable to
blending.

The radial velocity surveys have found that 0.5\%--1\% of Sun-like stars
harbour a Jupiter-mass companion in a 0.05 AU (3--5 day) orbit
\citep{lin03}. Since approximately 10\% of these planets with randomly
oriented orbits will transit the face of their parent star, a suitable
monitoring program of 20000 stars can be expected to yield $\sim 10$
transiting extra-solar planets. If we assume that the typical depth of
a transit signature is similar to that of the OGLE transiting planets
\citep{bou04,tor04}, then the photometric deviation for each transit
from a constant lightcurve will be $< 3$\%. Figure 5 shows that many of
the variables are of very low amplitude, with 20 having an rms of
$< 3$\%. Hence, for the Pegasus field, the number of low amplitude
variables detected outnumber the number of expected transits by a factor
of 2:1. As previously mentioned, the number of variables detected by the
WASP0 observations is relatively small compared to other similar
studies of field stars. Thus, it can be expected that transit detection
algorithms will suffer from large contamination effects due to variables,
unless additional stellar attributes, such as colour, can be incorporated
to reduce the false-alarm rate.

\section{Conclusions}

This paper has described observations of a field in Pegasus using the
Wide Angle Search for Planets prototype. These observations were
conducted from La Palma as part of a campaign to hunt for transiting
extra-solar planets. Careful monitoring of stars in the Pegasus field
detected 75 variable stars, 73 of which were previously unknown to be
variable. It is estimated from these observations that $\sim 0.4$\% of
Pegasus field stars brighter than $V \sim 13.5$ are variables with an
amplitude greater than $\sim 0.6$\%. This is relatively low compared to
other similar studies of field stars.

A concern for transiting extra-solar planet surveys is reducing the
number of false-alarms which often arise from grazing eclipsing binaries
and blended stars. It is estimated that the number of detectable
transiting extra-solar planets one could expect from the stars monitored
is a factor of two smaller than the number of variables with similar
photometric amplitudes. Since the number of variables detected is
relatively low, this has the potential to produce a strong false-alarm
rate from transit detection algorithms.

Surveys such as this one are significantly improving our knowledge of
stellar variability and the distribution of variable stars. It has
been a considerable source of interest for many to discover the extent
of sky that remains to be explored in this way, even to relatively
shallow magnitude depths. It is expected that our knowledge of variable
stars will continue to improve, particularly with instruments such as
SuperWASP \citep{str03b} intensively monitoring even larger areas of sky.

\section*{Acknowledgements}

The authors would like to thank Ron Hilditch for several useful
discussions. The authors would also like to thank PPARC for supporting
this research and the Nichol Trust for funding the WASP0 hardware.
This publication makes use of data products from the Two Micron All Sky
Survey, which is a joint project of the University of Massachusetts and
the Infrared Processing and Analysis Center/California Institute of
Technology, funded by the National Aeronautics and Space Administration
and the National Science Foundation.


\begin{thebibliography}{}
\bibitem[\protect\citeauthoryear{Alcock et al.}{2003}]{alc03} Alcock C.,
et al., 2003, ApJ, 598, 597
\bibitem[\protect\citeauthoryear{Bakos et al.}{2002}]{bak02} Bakos,
G.\'A., L\'az\'ar, J., Papp, I., S\'ari, P., Green, E.M., 2002, PASP,
114, 974
\bibitem[\protect\citeauthoryear{Bessell}{1990}]{bes90} Bessell, M.S.,
1990, PASP, 102, 1181
\bibitem[\protect\citeauthoryear{Borucki et al.}{2001}]{bor01} Borucki,
W.J., Caldwell, D., Koch, D.G., Webster, L.D., Jenkins, J.M., Ninkov, Z.,
Showen, R., 2001, PASP, 113, 439
\bibitem[\protect\citeauthoryear{Bouchy et al.}{2004}]{bou04} Bouchy, F.,
Pont, F., Santos, N.C., Melo, C., Mayor, M., Queloz, D., Udry, S., 2004,
A\&A, 421, L13
\bibitem[\protect\citeauthoryear{Brown}{2003}]{bro03} Brown, T.M., 2003,
ApJ, 593, L125
\bibitem[\protect\citeauthoryear{Delfosse et al.}{1998}]{del98} Delfosse,
X., Forveille, T., Mayor, M., Perrier, C., Naef, D., Queloz, D., 1998,
A\&A, 338, L67
\bibitem[\protect\citeauthoryear{Derue et al.}{2002}]{der02} Derue, F.,
et al., 2002, A\&A, 389, 149
\bibitem[\protect\citeauthoryear{Everett et al.}{2002}]{eve02} Everett,
M.E., Howell, S.B., van Belle, G.T., Ciardi, D.R., 2002, PASP, 114, 656
\bibitem[\protect\citeauthoryear{Hartman et al.}{2004}]{har04} Hartman,
J.D., Bakos, G,., Stanek, K.Z., Noyes, R.W., 2004, AJ, 128, 1761
\bibitem[\protect\citeauthoryear{Henry, Fekel, \& Henry}{Henry et
al.}{2004}]{hen04} Henry, G.W., Fekel, F.C., Henry, S.M., 2004, A\&A,
127, 1720
\bibitem[\protect\citeauthoryear{H{\o}g et al.}{2000}]{hog00} H{\o}g, E.,
et al., 2000, A\&A, 355, L27
\bibitem[\protect\citeauthoryear{Kane et al.}{2004}]{kan04} Kane, S.R.,
Collier Cameron, A., Horne, K., James, D., Lister, T.A., Pollacco, D.L.,
Street, R.A., Tsapras, Y., 2004, MNRAS, 353, 689
\bibitem[\protect\citeauthoryear{Kidger}{2004}]{kid04} Kidger, M.R., 2004,
AJ, submitted
\bibitem[\protect\citeauthoryear{Lineweaver \& Grether}{2003}]{lin03}
Lineweaver, C.H., Grether, D., 2003, ApJ, 598, 1350
\bibitem[\protect\citeauthoryear{Maceroni \& Montalb\'an}{2004}]{mac04}
Maceroni, C., Montalb\'an, J., 2004, A\&A, 426, 577
\bibitem[\protect\citeauthoryear{Mochejska et al.}{2002}]{moc02} Mochejska,
B.J., Stanek, K.Z., Sasselov, D.D., Szentgyorgyi, A.H., 2002, AJ, 123,
3460
\bibitem[\protect\citeauthoryear{Monet et al.}{2003}]{mon03} Monet, D.G.,
et al., 2003, ApJ, 125, 984
\bibitem[\protect\citeauthoryear{Ochsenbein, Bauer, \& Marcout}{Ochsenbein
et al.}{2000}]{och00} Ochsenbein, F., Bauer, P., Marcout, J., 2000, A\&AS,
143, 23
\bibitem[\protect\citeauthoryear{Press et al.}{1992}]{pre92} Press, W.H.,
Teukolsky, S.A., Vetterling, W.T., Flannery, B.P., 1992, Numerical Recipes
in FORTRAN: The Art of Scientific Computing (Cambridge University Press)
\bibitem[\protect\citeauthoryear{Ribas}{2003}]{rib03} Ribas, I., 2003,
A\&A, 398, 239
\bibitem[\protect\citeauthoryear{Rubenstein}{2001}]{rub01} Rubenstein,
E.P., 2001, AJ, 121, 3219
\bibitem[\protect\citeauthoryear{Street et al.}{2002}]{str02} Street, R.A.,
et al., 2002, MNRAS, 330, 737
\bibitem[\protect\citeauthoryear{Street et al.}{2003}]{str03a} Street, R.A.,
et al., 2003, MNRAS, 340, 1287
\bibitem[\protect\citeauthoryear{Street et al.}{2003}]{str03b} Street, R.A.,
et al., 2003, ASP Conf. Series, Vol. 294, Scientific Frontiers in Research
on Extrasolar Planets, eds. D. Deming \& S. Seager, p. 405
\bibitem[\protect\citeauthoryear{Torres et al.}{2004}]{tor04} Torres, G.,
Konacki, M., Sasselov, D.D., Jha, S., 2004, ApJ, 609, 1071
\bibitem[\protect\citeauthoryear{Wenger et al.}{2000}]{wen00} Wenger, M.,
et al., 2000, A\&AS, 143, 9
\bibitem[\protect\citeauthoryear{Wo\'zniak et al.}{2002}]{woz02} Wo\'zniak,
P.R., Udalski, A., Szyma\'nski, M., Kubiak, M., Pietrzy\'nski, G.,
Soszy\'nski, I., \.Zebru\'n, K., 2002, AcA, 52, 129
\bibitem[\protect\citeauthoryear{Wo\'zniak et al.}{2004}]{woz04} Wo\'zniak,
P.R., et al., 2004, AJ, 127, 2436
\end{thebibliography}
\end{document}